\documentclass[aps,prl,amsmath,amssymb,reprint,balancelastpage,floatfix,showkeys,showpacs,footinbib]{revtex4-1} 

\usepackage{graphicx}
\usepackage{subfigure}

\newcommand{\unit}[1]{\ensuremath{\, \mathrm{#1}}}

\begin{document}

\title{Rydberg atoms with a reduced sensitivity to dc and low-frequency electric fields}
\author{L.~A.~Jones}
\author{J.~D.~Carter}
\author{J.~D.~D.~Martin}
\affiliation{Department of Physics and Astronomy and Institute for Quantum Computing, University of Waterloo, Waterloo, Ontario, Canada N2L 3G1}
\date{\today}

\begin{abstract}
A non-resonant microwave dressing field at $38.465\unit{GHz}$ was used to eliminate the static electric dipole moment difference between the $49s_{1/2}$ and $48s_{1/2}$ Rydberg states of $^{87}$Rb in dc fields of $\approx 1 \unit{V/cm}$.  The reduced susceptibility to electric field fluctuations was measured using 2-photon microwave spectroscopy.   An anomalous spectral doublet is attributed to polarization ellipticity in the dressing field.  The demonstrated ability to inhibit static dipole moment differences --- while retaining sensitivity to high frequency fields --- is applicable to sensors and/or quantum devices using Rydberg atoms.
\end{abstract}

% temporarily comment this stuff out to get under 5 pages for arxiv submission
%\pacs{ %http://www.aip.org/pacs/pacs2010/individuals/pacs2010_regular_edition/index.html
%32.80.Ee, % Rydberg states
%32.10.Dk, % Electric and magnetic moments, polarizabilities
%32.60.+i % Zeeman and Stark effects
%}
%\keywords{Rydberg atoms, Stark effect, dressing fields}%Use showkeys class option if keyword 

\maketitle

Rydberg atoms have a high sensitivity to resonant oscillating electric fields, typically at GHz frequencies.  This property is useful for exploring the quantum nature of radiation-matter interactions \cite{raimond:2001}, and for high sensitivity rf/microwave detection \cite{bradley:2003,sedlacek:2012}.  But Rydberg atoms also have a potentially problematic sensitivity to lower-frequency non-resonant fields, which can destroy coherence. One source of spurious fields are nearby surfaces, as sensed by Rydberg atoms near metals \cite{sandoghdar:1996,*weidinger:1997,*pu:2010,*hattermann:2012}, co-planar resonators \cite{hogan:2011}, and both shielded \cite{tauschinsky:2010} and unshielded \cite{carter:2012} atom chips.   The low-frequency fields near surfaces and the high sensitivity of Rydberg atoms to these fields may limit schemes to couple these atoms to surface devices 
\cite{PhysRevLett.92.063601,petrosyan:2008,*petrosyan:2009,saffman:2010}.

It is often desirable for an atomic transition frequency to have a low sensitivity to electric and/or magnetic fields.  For example, the ``clock'' transitions between alkali hyperfine levels are insensitive to first-order magnetic field variations \cite{audoin:2001}.  When two-level systems are used to implement qubits, a reduced sensitivity to noisy perturbations helps preserve qubit coherence, as illustrated by Vion {\it et al.}~\cite{vion:2002} for superconducting qubits.

To reduce the influence of low-frequency electric fields on Rydberg atoms Hyafil {\it et al.}~\cite{hyafil:2004,*mozley:2005} have proposed using non-resonant dressing microwave fields to modify electric field susceptibilities.  A dressing field may be chosen to minimize the dc electric field sensitivity of a transition frequency between two high-angular momentum ``circular'' Rydberg states while retaining their sensitivity to higher frequency fields.   Dressing fields are a useful tool for the modification of atomic properties: they can reduce the influence of magnetic fields on optical clock transitions \cite{zanon:2012}, enhance interatomic Rydberg interactions \cite{bohlouli:2007,*petrus:2008,*tauschinsky:2008,*brekke:2012,*tanasittikosol:2011}, and can increase Rydberg susceptibility to electric fields \cite{bason:2010}.

In this work we experimentally demonstrate {\em dipole nulling} -- application of a microwave dressing field to dramatically reduce Rydberg atom susceptibility to varying dc electric fields -- using low-angular momentum states.  Specifically we study the 2-photon $49s_{1/2} \rightarrow 48s_{1/2}$ transition of $^{87}$Rb in a field of $1\unit{V/cm}$ and find that we can eliminate the first-order dependence on fluctuations about this field.
Rydberg atoms may require a dc electric field to break degeneracies \cite{raimond:2001}, enhance interatomic interactions by making them resonant \cite{safinya:1981,gallagher:1994} or for trapping \cite{hyafil:2004}.  Surprisingly, we find that a dressing field can make the transition energy less sensitive to dc electric field fluctuations around $1\unit{V/cm}$ than at zero dc field with no dressing field.  This is despite the fact that low angular momentum Rydberg states are ordinarily more sensitive to field fluctuations in the presence of a dc field.

The influence of dressing fields may be considered using exact diagonalization of truncated Floquet matrices \cite{shirley:1965,gallagher:1994}.
However, perturbation theory aids the choice of an appropriate dressing frequency and power for nulling.  We use the methods of Zimmerman {\it et al.}~\cite{zimmerman:1979} to construct and diagonalize a Hamiltonian in a non-zero dc field $F_{dc,0}$ with no dressing field.  The oscillating field (of amplitude $F_{ac}$) and deviations in the static dc field $\Delta F_{dc}=F_{dc}-F_{dc,0}$ are then considered as perturbations to the $ith$ state, shifting the unperturbed energy $E_{i,0}$:
\begin{equation}
E_i\approx E_{i,0}-\mu_i\Delta F_{dc}-\tfrac{1}{4}\alpha_{i}(\omega)F_{ac}^2+\beta_i(\omega) F_{ac}^2 \Delta F_{dc},
\label{eq:perturbation_expansion}
\end{equation}
where $\mu_i$ is the undressed $F_{dc,0}$ induced dipole moment, 
$\alpha_{i}(\omega)$ is the ac polarizability, and $-\beta_i(\omega) F_{ac}^2$ is an ac field induced contribution to the dipole moment (calculated using 3{rd} order perturbation theory). Deviations in the dc field $\Delta F_{dc}$ are considered to be in the same direction as $F_{dc,0}$ as these cause the first-order shifts that we would like to suppress, whereas symmetry dictates that fluctuations {\em transverse} to $F_{dc,0}$ only result in 2nd order (and higher) shifts \footnote{We neglect the 2nd order term that is quadratic in $\Delta F_{dc}$, which can be understood as a slight perturbation of the dc field-induced dipole moment due to the presence of $\Delta F_{dc}$. This perturbation is assumed to be small compared to $\mu_i$ and $\beta_i(\omega) F_{ac}^2$.  In our comparison between experiment and theory we establish the limits of this perturbation theory result by diagonalization of Floquet Hamiltonians.}.  
We consider a dressing field applied in the same direction as the static field, as boundary conditions near metal surfaces dictate that this is the most likely scenario to be encountered in practice.

In the absence of the dressing field, the difference in the dipole moments between two states $\Delta \mu = \mu_{1}-\mu_{2}$ dictates the first-order dependence of the transition frequency on fluctuations in the dc field.  The dressing field induced contribution to the dipole moment can be used to counteract this sensitivity:  setting $-\Delta \mu + \Delta \beta(\omega)F_{ac}^2=0 $ eliminates the first-order dependence of the transition frequency to field variations $\Delta F_{dc}$ about $F_{dc,0}$.  

\begin{figure}
\includegraphics[width=3.375in]{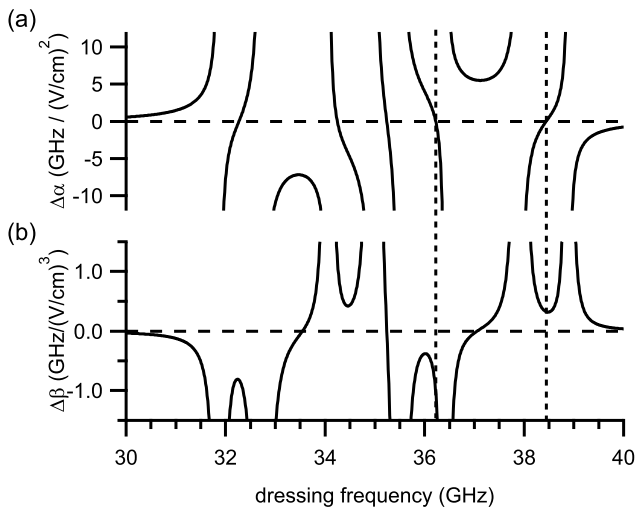}
\caption{
a) ac polarizability, and b) dressing field induced dipole moment differences (see Eq.~\ref{eq:perturbation_expansion}) between the $49s_{1/2}$ and $48s_{1/2}$ states of $^{87}$Rb in a $1\unit{V/cm}$ dc field.
\label{fg:alphabeta}}
\end{figure}

To demonstrate dipole nulling we have chosen the $49s_{1/2} \rightarrow 48s_{1/2}$ transition of $^{87}$Rb in a dc field of $1\unit{V/cm}$.  This choice is based on practical considerations:  given a $(n+1)s_{1/2} \rightarrow ns_{1/2}$  system in Rb, this is the lowest $n$ for which the involved frequencies (probe and dressing) \cite{li:2003} are below the $40\unit{GHz}$ upper frequency limit of our available synthesizers \footnote{Initial attempts to demonstrate dipole nulling using the $36s_{1/2}-37s_{1/2}$ levels were unsuccessful due to excessive broadband spectral noise in our dressing field source (an active frequency multiplier) at $\approx 98 \unit{GHz}$, which drove unwanted resonant single-photon transitions (e.g. $36s_{1/2}-35p_{1/2,3/2}$).}.
Figure \ref{fg:alphabeta}b illustrates the results of a perturbative calculation of $\Delta \beta(\omega)=\beta_{49s_{1/2}}(\omega)-\beta_{48s_{1/2}}(\omega)$ for these two states.  As  $\Delta \mu = \mu_{49s_{1/2}}-\mu_{48s_{1/2}}$ 
is positive ($6.0\unit{MHz/(V/cm)}$) we are confined to dressing frequencies $\omega$ where $\Delta \beta(\omega)$ is positive, to satisfy $-\Delta \mu + \Delta \beta(\omega)F_{ac}^2=0$.  Figure \ref{fg:alphabeta} shows two frequency ranges where $\Delta \beta (\omega) > 0$.

\begin{figure}
\includegraphics[width=3.375in]{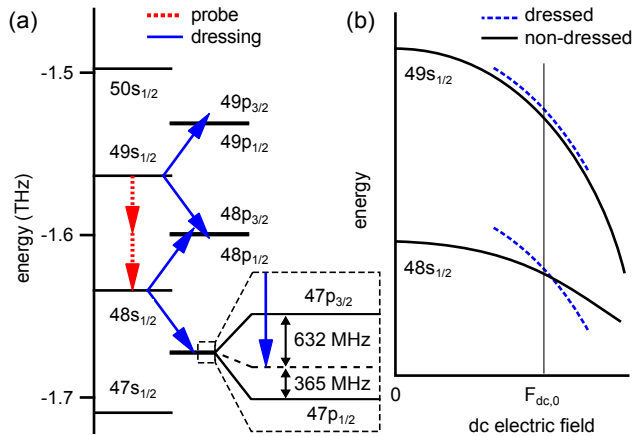}
\caption{
(a) Zero field atomic energy levels and the microwave dressing and probe frequencies. 
(b) The dipole nulling effect under ideal conditions:
with the application of a dressing field there is no first-order dependence of the transition energy on the dc electric field and no differential ac Stark shift at $F_{dc,0}$.
}
\label{fg:dressednondressed}
\end{figure}

The remaining flexibility in the choice of $\omega$ may be used to satisfy a second constraint (analogous to the magic wavelengths for optical Stark shifts \cite{ye:2008}).  By choosing the difference in the ac polarizabilities $\Delta \alpha(\omega)$ to be zero, we can reduce the influence of inhomogeneities in the dressing field over the sample \footnote{
Satisfying $\Delta \alpha(\omega)=0$ ensures that inhomogeneities in the dressing field do not contribute to differential energy levels shifts --- inhomogeneities will only influence the efficacy of the dipole nulling.}.  The condition $\Delta \alpha (\omega)= 0$ occurs at a frequency of $38.44\unit{GHz}$ (see Fig.~\ref{fg:alphabeta}), where $\Delta \beta =0.37\unit{GHz/(V/cm)^3}$ and thus for nulling $F_{ac} = \sqrt{\Delta \mu / \Delta \beta} = 0.13\unit{V/cm}$.  This is a relatively weak microwave field amplitude -- an enhancement cavity is not necessary.  
Figure \ref{fg:dressednondressed} illustrates the relevant energy levels for the particular $n$ we have studied.

\begin{figure}
\centering
\includegraphics[width=3.375in]{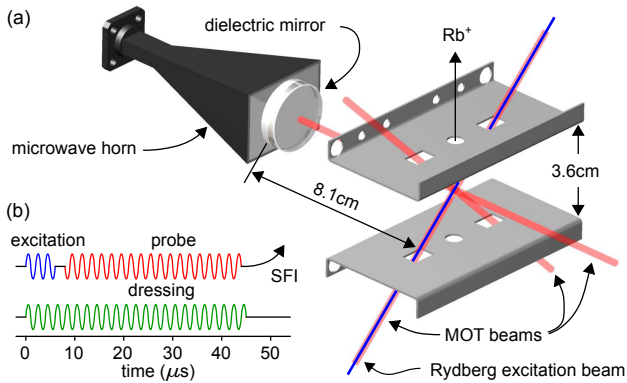}
\caption{
(a) Experimental apparatus.
(b) The sequence of applied fields. Prior to Rydberg excitation, the anti-Helmholtz coil current is ramped down and the residual magnetic field nulled using compensation coils; the coil current is switched on again once selective field ionization (SFI) is complete.  This sequence is repeated at $10\unit{Hz}$ as the probe frequency is stepped, and the Rydberg populations measured using SFI.
}
\label{fg:apparatus}
\end{figure}

Rubidium atoms were gathered in a $^{87}$Rb magneto-optical trap (MOT) and optically excited from the $5p_{3/2}, F=3$ state to the $49s_{1/2}, F=2$ Rydberg level using a frequency doubled Ti:sapphire laser, stabilized using a transfer cavity \cite{cels2:2011}.  DC electric fields were applied using two field plates positioned above and below the MOT center (see Fig.~\ref{fg:apparatus}a). The dressing microwave field was on continuously during optical excitation, while the energy difference between Rydberg states was probed using a second microwave field after optical excitation (see Fig.~\ref{fg:apparatus}b).  Rydberg state populations as a function of probe frequency were measured using selective field ionization (SFI) \cite{gallagher:1994}.  The strength of the probe was chosen to avoid excessive power broadening of the spectral line.

The dressing field was applied through a retro-reflecting mirror and vacuum window (see Fig.~\ref{fg:apparatus}a).  This configuration was chosen to reduce reflections from surfaces within the vacuum chamber, as these reflections cause standing waves which decrease the homogeneity of the dressing field.  Likewise the Rydberg excitation beam was orthogonal to the microwave propagation direction to reduce the influence of partial standing waves along the direction of microwave propagation.

\begin{figure}
\includegraphics[width=3.375in]{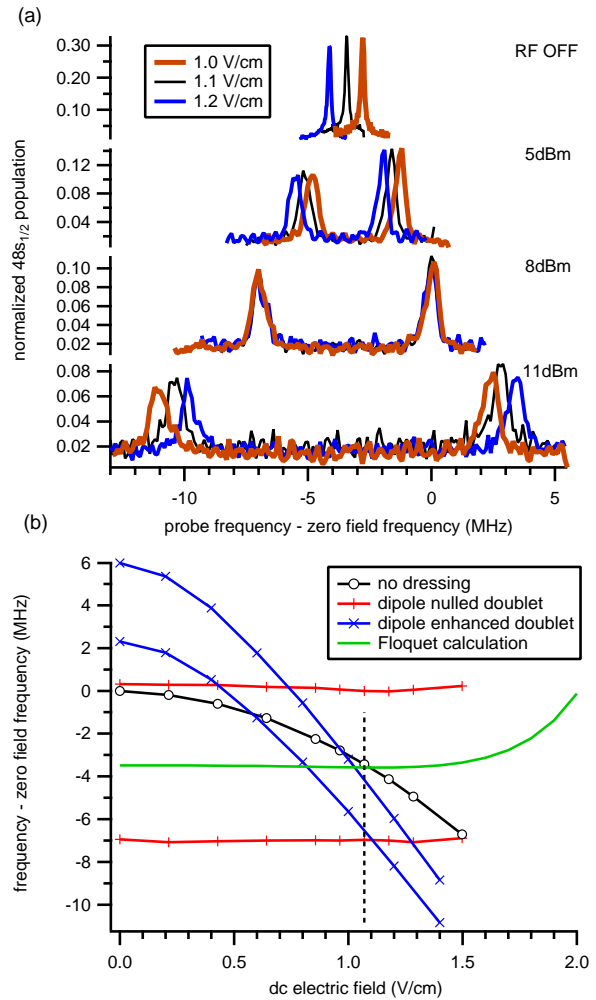}
\caption{
(a) The $49s_{1/2}\rightarrow48s_{1/2}$ probe transition for different microwave dressing powers and dc electric fields (labeled).  At $8\unit{dBm}$ dressing power  the spectral lines for the three different dc fields coincide, illustrating the dipole nulling effect.   
The dressing frequency is $38.465 \unit{GHz}$, chosen so that the doublet has zero average ac Stark shift at $1.1\unit{V/cm}$. 
% This dressing frequency is slightly different from the calculated $\Delta \alpha = 0$ frequency of $38.44\unit{GHz}$.
(b) Observed frequencies of the $49s_{1/2}-48s_{1/2}$ probe transitions as a function of dc electric field under different dressing field conditions, illustrating the dipole nulling and enhancement effects, compared with the non-dressed case. Estimated statistical errors are on the order of the symbol sizes.  Also shown are the results of a Floquet diagonalization with a linearly polarized dressing field ($F_{ac}=0.13 \unit{V/cm}$ and $\omega/(2\pi) = 38.465\unit{GHz}$).  Since the probe is a two-photon transition, ``frequency'' in both (a) and (b) refers to twice the applied frequency, for correspondence with energy level differences.
}
\label{fg:nullingexperiment}
\end{figure}

Figure \ref{fg:nullingexperiment} shows probe spectra of the $49s_{1/2}-48s_{1/2}$ transition with different dc fields and varying dressing powers (the powers quoted are the synthesizer level settings).  As the dressing power is increased the probe line splits into a doublet, which we attribute to a polarization imperfection in the dressing field (discussed further below).  Nulling occurs for both lines of the doublet.  In particular, we can compare the non-dressed to dressed Rydberg state transition frequencies at several dc fields as shown in Fig.~\ref{fg:nullingexperiment}b. There is a distinct dipole moment at $1\unit{V/cm}$ in the absence of dressing, whereas the first-order dependence on dc field is eliminated when the dressing microwave field is present.  The nine measured peak centers of the dipole-nulled lower sideband have a standard deviation of $0.06\unit{MHz}$ over the dc field range $0$ to $1.5\unit{V/cm}$ compared with the systematic deviation of $7\unit{MHz}$ over the same range when no dressing field is applied.  The strength of the two-photon probe coupling remains unchanged, with slight line broadening due to dressing field inhomogeneity.

Both the observations and the full Floquet diagonalization show surprisingly small variations in the dressed energies with dc field -- even down to low fields.  In fact, $\Delta \beta$ is very nearly proportional to $F_{dc,0}$ together with $\Delta \mu$. This suggests an alternative theoretical formulation where we consider dressing field induced changes in the polarizabilities about zero dc field (rather than induced dipole moments about non-zero fields).  The perturbations to achieve ``polarizability nulling'' would be 4th order (2nd order in each of the fields). However, this alternative approach will break down when considering coupling to states in the Stark manifold whose dipole moments are independent of dc field.  In this case only first order suppression is possible. The dipole nulling description --- making use of perturbation expansions around non-zero dc fields --- is a more general approach.

Bason {\it et al.}~\cite{bason:2010} have demonstrated that non-resonant dressing fields can {\em enhance} Rydberg susceptibility to dc electric fields.  In our case it is possible to increase the dipole moment difference between the $49s_{1/2}$ and $48s_{1/2}$ states by using a dressing frequency where $\Delta \beta (\omega)$ is the opposite sign of $\Delta \mu$.  Figure \ref{fg:nullingexperiment}b shows that the $49s_{1/2}-48s_{1/2}$ transition frequency becomes more sensitive to dc field fluctuations when a dressing frequency of $36.225\unit{GHz}$ is applied, consistent with the calculation of $\Delta \beta (\omega)$ in Fig.~\ref{fg:alphabeta}; i.e. $\Delta \beta (\omega) < 0$ for $\omega/(2\pi) = 36.225\unit{GHz}$.  However, application of a non-zero dc bias electric field --- in situations where this is possible --- is a more straightforward way to increase Rydberg susceptibility to electric fields.

We now discuss the origin of the doublet in the probe spectra.  
This is not an Autler-Townes doublet, as the dressing frequency is detuned from the nearest transition by 365 MHz  (see inset to Fig.~\ref{fg:dressednondressed}), which is much larger than the observed splitting.  This splitting also occurs at zero electric field and does not depend on Rydberg atom density. As we show below, by varying the microwave polarization we have established that this splitting is due to a slight polarization ellipticity in the dressing field.  
Although the microwaves emerging from the horn should have a high degree of polarization purity, reflections from the vacuum chamber walls, field plates, and windows can introduce ellipticity in the field at the location of atoms.   
The splitting is similar to the magnetic field-like vector ac Stark shift observed for ground-state atoms in circularly polarized fields \cite{happer:1967}.

To understand the splitting, it is useful to consider two extreme cases: 1) a linearly polarized dressing field, and 2) a circularly polarized dressing field.  In the linearly polarized field, with an axis of quantization along the oscillating field direction, the $m_j=\pm 1/2$ states are uncoupled and show the same ac Stark shifts.  In a purely circular field, a natural axis of quantization is normal to the plane containing the oscillating E field.  With this choice, the $m_j=\pm1/2$ states are also uncoupled, but show different ac Stark shifts, due to their different allowed couplings (i.e. with $\sigma_{+}$: $s_{1/2},m_j=1/2$ is coupled to $p_{3/2},m_j=3/2$ whereas $s_{1/2},m_j=-1/2$ is coupled to $p_{1/2},m_j=1/2$
and $p_{3/2},m_j=1/2$).  Elliptical polarizations are intermediate between these two extremes, and the splitting can be calculated using Floquet theory.  Although both the $49s_{1/2}$ and $48s_{1/2}$ states split, the calculations indicate that splitting of the $49s_{1/2}$ state is 40 times smaller than that of the $48s_{1/2}$ state and thus unresolved in our probe spectra \footnote{The second doublet is unresolvable in the spectra of Fig.~\ref{fg:nullingexperiment}: at the largest dressing field the splitting is predicted to be $350\unit{kHz}$, whereas the linewidth is 
$800\unit{kHz}$.  As both the splitting and the broadening due to field inhomogeneities scale linearly with dressing power, the $49s_{1/2}$ splitting could only be resolved by improving the dressing field homogeneity.}.

\begin{figure}
\centering
\subfigure{
\raisebox{1.in}{\hbox{(a)}}
\includegraphics[width=3.375in]{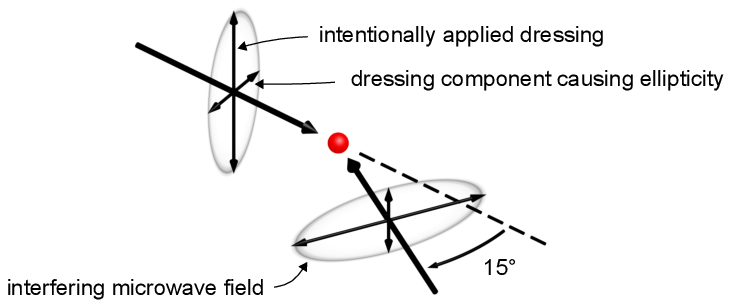}} \\
\subfigure{\raisebox{2.05in}{\hbox{(b)}} % \subfigure{\raisebox{2.3in}{\hbox{(b)}}
\includegraphics[width=3.375in]{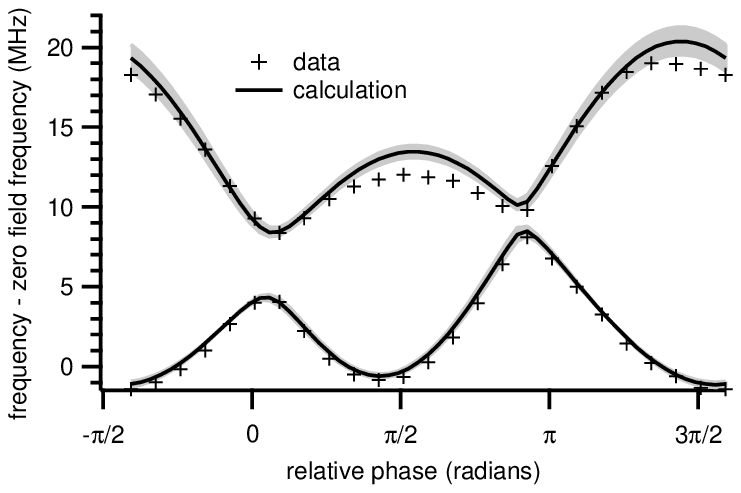}}
\caption{
(a) Arrangement of interfering (nearly) orthogonal microwave polarizations for verifying splitting is due to ellipticity.
(b) The location of two lines of the $48s_{1/2}-49s_{1/2}$ doublet as a function of phase between two interfering dressing fields at $38.1\unit{GHz}$.  The experimental data is shifted horizontally by a constant phase for best agreement with the calculation.  The shading around the calculation indicates uncertainty related to variations in the transmitted amplitude through the phase shifter.  Uncertainties in the measured line positions are on the order of the symbol sizes.
}
\label{fg:ellipticity}
\end{figure}

To confirm that polarization is responsible for the splitting we aimed a second horn towards the atoms (see Fig.~\ref{fg:ellipticity}a).  The microwave dressing signal from a synthesizer was split and sent to each of the two horns, with adjustable relative phase and power.  The intensity and ellipticity of the fields due to each horn was determined by observing the splitting of the $49s_{1/2}-48s_{1/2}$ probe transition, using a dressing field frequency where the ac Stark shift is non-zero ($38.1\unit{GHz}$ -- see Fig.~\ref{fg:alphabeta}) and selectively blocking one or the other of the horns.  The field intensities were determined from the observed average ac Stark shift and the theoretical $\Delta \alpha$. The ellipticities were determined by comparison of the ratio of the splitting to the ac Stark shift (both scale linearly with microwave power), with a Floquet calculation. We characterize the polarization ellipses by the ratio of the minor to major axis: $-17\unit{dB}$ for one of the horns and $-11\unit{dB}$ for the other.

The relative powers were adjusted so that the intensity at the location of the atoms was the same for both horns, and then both beams were allowed to interfere. Figure \ref{fg:ellipticity}b shows that as the relative phase between the two interfering fields is changed --- modulating the polarization --- the splitting varies in magnitude. We can compute the expected splitting as a function of phase between the two fields using Floquet theory (see Fig.~\ref{fg:ellipticity}b).  The relative orientation of the two polarization ellipses is varied for best agreement with the data.  The discrepancies ($< 3 \unit{MHz}$) are possibly due to variation of the relative phase over the sample, and/or variations in transmitted power through the phase shifter with varying phase shift.  

Since the observed splitting is associated with the $m_j=\pm 1/2$ degeneracy of the $ns_{1/2}$ states, this degeneracy could also be lifted in a more controllable way by application of a dc magnetic field.  In this case, an elliptical dressing field would cause no further splitting, and the dressing frequency could be tuned to avoid any differential ac Stark shift for the component of interest.

Our choice of $n$ for dipole-nulling was based on available equipment, but nulling could be achieved for many Rb $(n+1)s_{1/2}-ns_{1/2}$ pairs.  In particular, the dressing frequency (for $\Delta \alpha (\omega) =0$) scales like $1/n^{3}$, and the ac-field magnitude to obtain zero dc quadratic Stark shifts about $F_{dc}=0$ scales like $F_{ac} \propto 1/n^6$ (determined by diagonalization of Floquet Hamiltonians for $n=30$ to $55$). Nulling with different species and transitions can be examined using the perturbative approach presented here (of particular interest are low-angular momentum states connected by a single-photon transition \cite{hogan:2011}).

In summary, we have demonstrated that it is possible to selectively inhibit the dipole moment difference between two Rydberg states using non-resonant dressing fields.  This technique offers a means to preserve the coherence of Rydberg atom superpositions and is complementary to spin-echo/refocussing techniques \cite{minns:2006,*yoshida:2008}.
It will be useful in situations where spatially inhomogeneous fields and/or low frequency fluctuating fields are present, such as near surfaces \cite{sandoghdar:1996,*pu:2010,*hattermann:2012,hogan:2011,tauschinsky:2010,carter:2012,carter:2011,muller:2011}.  Dressing fields could help maintain resonance between Rydberg atoms and high-Q devices on surfaces --- such as superconducting resonators \cite{PhysRevLett.92.063601} --- in the presence of uncontrolled low-frequency electric fields.  

We thank T.~F.~Gallagher and S.~Safavi-Naeini for useful discussions, and W.-K.~Liu and S. Graham for comments on this manuscript.  This work was supported by NSERC.  

\end{document}